\documentstyle{mn}
\input epsf.sty

\newif\ifAMStwofonts
\AMStwofontstrue
 
\ifoldfss
  \ifCUPmtlplainloaded \else
    \NewTextAlphabet{textbfit} {cmbxti10} {}
    \NewTextAlphabet{textbfss} {cmssbx10} {}
    \NewMathAlphabet{mathbfit} {cmbxti10} {} 
    \NewMathAlphabet{mathbfss} {cmssbx10} {} 
  \fi
  \ifAMStwofonts
    \ifCUPmtlplainloaded \else
      \NewSymbolFont{upmath} {eurm10}
      \NewSymbolFont{AMSa} {msam10}
      \NewMathSymbol{\upi}     {0}{upmath}{19}
      \NewMathSymbol{\umu}     {0}{upmath}{16}
      \NewMathSymbol{\upartial}{0}{upmath}{40}
      \NewMathSymbol{\leqslant}{3}{AMSa}{36}
      \NewMathSymbol{\geqslant}{3}{AMSa}{3E}

       \let\le=\leqslant
       
    \fi
  \fi
\fi 
 
\ifnfssone
  \newmathalphabet{\mathit}
  \addtoversion{normal}{\mahit}{cmr}{m}{it}
  \addtoversion{bold}{\mathit}{cmr}{bx}{it}
  \newmathalphabet{\mathbfit} 
  \addtoversion{normal}{\mathbfit}{cmr}{bx}{it} 
  \addtoversion{bold}{\mathbfit}{cmr}{bx}{it}
  \newmathalphabet{\mathbfss} 
  \addtoversion{normal}{\mathbfss}{cmss}{bx}{n}
  \addtoversion{bold}{\mathbfss}{cmss}{bx}{n}
  \ifAMStwofonts
    \ifCUPmtlplainloaded \else
and
your
      \UseAMStwoboldmath
      \makeatletter
      \new@mathgroup\upmath@group
      \define@mathgroup\mv@normal\upmath@group{eur}{m}{n}
      \define@mathgroup\mv@bold\upmath@group{eur}{b}{n}
      \edef\UPM{\hexnumber\upmath@group}
      \new@mathgroup\amsa@group
      \define@mathgroup\mv@normal\amsa@group{msa}{m}{n}
      \define@mathgroup\mv@bold\amsa@group{msa}{m}{n}
      \edef\AMSa{\hexnumber\amsa@group}  
      \makeatother
      \mathchardef\upi="0\UPM19
      \mathchardef\umu="0\UPM16
      \mathchardef\upartial="0\UPM40
      \mathchardef\leqslant="3\AMSa36
      \mathchardef\geqslant="3\AMSa3E

       \let\le=\leqslant

    \fi
  \fi
\fi 
   
\ifnfsstwo
  \DeclareMathAlphabet{\mathbfit}{OT1}{cmr}{bx}{it}
  \SetMathAlphabet\mathbfit{bold}{OT1}{cmr}{bx}{it}
  \DeclareMathAlphabet{\mathbfss}{OT1}{cmss}{bx}{n}
  \SetMathAlphabet\mathbfss{bold}{OT1}{cmss}{bx}{n}
  \ifAMStwofonts
    \ifCUPmtlplainloaded \else
      \DeclareSymbolFont{UPM}{U}{eur}{m}{n}
      \SetSymbolFont{UPM}{bold}{U}{eur}{b}{n}
      \DeclareSymbolFont{AMSa}{U}{msa}{m}{n}
      \DeclareMathSymbol{\upi}{0}{UPM}{"19}
      \DeclareMathSymbol{\umu}{0}{UPM}{"16}
      \DeclareMathSymbol{\upartial}{0}{UPM}{"40}
      \DeclareMathSymbol{\leqslant}{3}{AMSa}{"36}
      \DeclareMathSymbol{\geqslant}{3}{AMSa}{"3E}

       \let\le=\leqslant

    \fi
  \fi  
\fi 
      
\ifCUPmtlplainloaded \else
  \ifAMStwofonts \else 
    \def\upi{\pi}
    \def\umu{\mu}
    \def\upartial{\partial}
  \fi
\fi
                  
\title{Vertical Structure of Accretion Discs with Hot Coronae in AGN}
\author[A. R\' o\. za\' nska, B. Czerny, P.T. \. Zycki, G. Pojma\'nski]
       {A. R\' o\. za\' nska$^1$, B. Czerny$^1$, P.T. \. Zycki$^{1,2}$, 
        G. Pojma\'nski$^3$\\
        $^1$ Nicolaus Copernicus Astronomical Center, Bartycka 18, 00-716 Warsaw, 
        Poland \\
        $^2$ Department of Physics,University of Durham, South Road, Durham 
DH1 3LE, England \\
        $^3$ Astronomical Observatory of Warsaw University, Aleje 
Ujazdowskie 4,00-478 Warsaw, Poland } 

\begin{document}

\maketitle

\begin{abstract}

We study vertical structure of radiation pressure dominated 
disc with a hot corona. We include all the relevant processes like
bound--free opacity and convection. We show that the presence of the corona
modifies considerably the density and the opacity of the disc surface layers
which are important from the point of view of spectrum formation. The 
surface of the disc with a corona is much denser and less ionized than the
surface of a bare disc. Such a disc is likely to produce a neutral 
reflection 
and a local spectrum close to a black body. This effect will help to 
reconcile the predictions
of accretion disc models with the observational data since a neutral 
reflection
and a lack of Lyman edge are generally seen in AGN.

\end{abstract}


\section{Introduction}

Although it is widely believed that accretion onto supermassive black hole 
is the ultimate source of power in AGN we are still far from understanding
the details of this process.

Recent multiwavelenth observations clearly show that the gas in the vicinity
of the center forms a complex multi-phase medium. Cold gas is seen due to its
imprints on the reflected X-ray radiation (Pounds et al. 1990)
whilst hot plasma is responsible for the hard X-ray emission (for a review, see
Mushotzky, Done \& Pounds 1993).  The cold medium is also 
thought to be responsible 
for the optical/UV/Soft X-ray emission, i.e. big bump. 

The geometrical 
arrangement of the two phases is not clear but the most attractive model 
consists of an accretion disc surrounded by a hot corona, with the energy
being released in both phases (e.g. Edelson et al. 1997 for NGC 4151). 
The disc in some objects 
may possibly
be disrupted in its innermost part (e.g. $\sim  10 R_{Schw}$ for NGC 5548;
Loska \& Czerny 1997) and the flow may be dominated by advection (e.g. Narayan, 
Kato \& Tonma 1997).

There is, however, one problem which seems to be difficult to accommodate within
that scenario and which served frequently as an argument against the very existence 
of accretion discs in AGN. Namely, all realistic disc models predicted strong
atomic features, including strong Lyman edge whilst no such feature was clearly
detected in most of the AGN spectra (Antonucci 1993; see also 
Blaes \& Agol 1996). The problem cannot be overcome
by assumption that the disc surface is highly ionized and therefore no Lyman edge is
expected because the X-ray reflection puts a clear limit on the ionization 
parameter $\xi$
for a number of Seyfert 1 galaxies of order of 200 (\. Zycki et al. 1994). A huge Lyman
edge is seen from the gas irradiated by X-rays with such a value of  $\xi$, and
it is still clearly seen if $\xi$ is an order of magnitude higher (e.g. 
Collin-Souffrin et al. 1996). The Lyman edge disappears completely only if the
temperature of the gas is above 1 million K or $\xi$ is above 3000 which is clearly
excluded by X-ray data.

In this paper we show that this apparent problem disappears if the boundary
conditions between the disc and the corona are described appropriately. The basic
effect is the hydrostatic pressure exerted on the disc surface by the corona
which leads to much higher densities at the disc surface then adopted in typical
computations which forces the local emission of the surface to be  
close to a black body.
We also study the effect of the corona on the disc interior.

\section{The model}

\subsection{Global disc parameters}

 The structure of the corona depends on the total  flux, 
$F_d + F_c$, generated in the disc and in the corona,
and the local Keplerian velocity, $\Omega_K$. Both values can be determined
at a given radius $r$ from the mass of the black hole, $M$, and the accretion 
rate $\dot M$ through the usual relations:
\begin{equation}
\Omega_K = \left({G M \over r^3}\right)^{1/2}
\end{equation}
and
\begin{equation}
F_d + F_c = {3 G M \dot M \over 8 \pi r^3} f(r)
\end{equation}
where $f(r)$ represents the boundary condition at the mar\-gi\-nal\-ly stable 
orbit
\begin{equation}
f(r)=1-(3R_{\rm Sch}/r)^{1/2}
\end{equation}
in the Newtonian approximation.

The value of the ratio $f$ given by the formula:
\begin{equation}
f=\frac {F_c}{F_d + F_c}
\end{equation}
is the free parameter in the model.

We frequently use dimensionless accretion rate,
\begin{equation}
\dot m \equiv{\dot M\over \dot M_{Edd}},
\end{equation}
where $\dot M_{Edd}$ is the critical (Eddington) accretion rate 
\begin{equation}
\dot M_{Edd} = {L_{Edd} \over c^2 \eta}={4\pi G M m_H \over \sigma_T c \eta},
\end{equation}
assuming the efficiency of accretion $\eta= 1/12$, as it results from the 
Newtonian approximation to disc accretion.

\subsection{The vertical structure of the disc}

\subsubsection{Local viscosity model}

The physics of accretion is poorly understood as the microscopic mechanism of
the angular momentum transfer remains unknown. Most promising, perhaps,
is the viscosity provided by the small scale magnetic field structure
which develops in the disc at the expense of its rotational energy.
The results of the computer simulations of this process 
(Balbus \& Hawley 1991, 1992,
Hawley, Gammie \& Balbus 1995; see also Balbus \& Hawley 1998 for a 
review) are definitely encouraging.

Therefore, as in the case of stellar convection, we restore to a
replacement of all the missing physics by a single parameter
$\alpha$ introduced by Shakura \& Sunyaev (1973). This kind of parameterization
is very convenient. What is more, some physical attempts to describe the global
effect of viscosity can be reduced to this scaling (e.g. Tout \& Pringle 1992,
Canuto, Goldman \& Hubickyj 1984, Hawley et al. 1995).

This assumption was widely used to calculate the vertically averaged
disc structure in AGN (e.g. Ross, Fabian \& Mineshige 1993; see also Frank, 
King \& Raine 1985).

It was also successfully used to calculate the vertical structure of accretion 
discs in cataclysmic variables leading to the discovery of the nature of
dwarf novae eruptions (Meyer \& Meyer-Hofmeister 1981 and Smak 1982b).
The adoption of the viscosity $\alpha$ locally allows to calculate the local
behaviour of the disc. Thus the models include realistic opacities and 
the energy transport
by convection (see Cannizzo 1994 for a recent review). There were also 
two-dimensional studies of models
which included a complex inflow pattern (Urpin 1984, Siemiginowska 1988,
Kley \& Lin 1992, R\' o\. zyczka, Bodenheimer \& Bell 1994), consisting,
for low viscosity, of outflow near equatorial plane and inflow close to the 
surface, but resulting in  pure inflow for high viscosity ($\alpha > 0.1$). 

These results for cataclysmic variables cannot be easily adopted to the case 
of AGN accretion discs.
In accretion discs in cataclysmic variables the radiation pressure is always 
negligible  due to
the presence of an extended central body.

However, certain progress has been made also in AGN discs.

The local disc structure was initially studied under an assumption that the 
stress
is proportional to the gas pressure (not to the total pressure)  (e.g. Lin \&
Shields 1986, Mineshige \& Shields 1990). Those studies concentrated on 
conditions in which the ionization instability may
operate.
Shimura \& Takahara (1993) studied
the disc structure under the assumption
that the energy generation rate per unit volume is proportional to the density.
Outermost parts of the disc (mostly dominated by 
the gas pressure anyway) were studied with the $\alpha$--prescription 
e.g.\ by Cannizzo (1992);
Cannizzo \& Reiff (1992) and Hur\'{e} et al. (1994a,b).

The computations of the vertical structure of the inner radiation pressure 
dominated parts of the disc under the assumption of the viscosity proportion
 to the total (i.e. gas plus radiation) pressure were postponed for a long time
because of the problem of instability discovered by Pringle, Rees \&
Pacholczyk (1973) and discussed by Shakura \& Sunyaev (1976). It was 
initially suggested that the disc in that region of the size of a 
thousand Schwarzschild radii is replaced by the hot
optically thin medium (Shapiro, Lightma \& Eardley 1976). However, direct
observational evidences show the presence of the cold medium very  close to 
the black hole (see Mushotzky et al. 1993 for a review and Mushotzky et al. 
1995, Tanaka et al. 1995, Yaqoob et al. 1996 and a summary by Nandra et al. 
1997
for the evidences of the cold disc based on the iron $K_{\alpha}$ line).
The theoretical reason for the disc survival is not clear although recently
more advanced studies of the disc stability was undertaken (Gammie 1998). 
Since we 
do not know whether the disc survives due to the stabilizing role of the 
corona or the disruption of the disc leads to cold clump formation with the 
clumps themselves simulating the behaviour of the turbulent disc (e.g.
Collin-Souffrin et al. 1996, Krolik 1998) one of the possible approaches is
to explore the possible models irrespective of the potential stability 
problems and to verify them observationally.

Recently,  the vertical structure of a radiation-pressure 
dominated accretion disc around a massive black hole was studied 
by Milson et al. (1994) and D\" orrer et al. (1996). 
We follow their general line apart from the condition 
on the
disc surface  which are imposed by the presence of the corona, and some other
minor modifications.

\subsubsection{Equations}

We assume the standard viscosity model of Shakura \& Sunyaev (1973) with the
viscosity proportional to the total (i.e. gas plus radiation) pressure.

We compute the vertical structure using standard equations
modified  by the presence of convection (Shakura \& Sunyaev 1976;
Meyer \& Meyer-Hofmeister 1981):
\begin{equation}
{1 \over \rho} {dP \over dz} = - \Omega_K^2 z,  
\end{equation}
\begin{equation}
P = P_{gas} + P_{rad},
\end{equation}
\begin{equation}
F=F_{\rm rad}= -{16 \sigma T^3  \over 3\kappa \rho }{dT \over dz},
\ \ \ \nabla_{\rm rad}\le\nabla_{\rm ad}, 
\end{equation}
\begin{equation}
F = F_{\rm rad} + F_{\rm conv}, \ \ \ \ \ \nabla_{\rm rad}>\nabla_{\rm ad}.
\end{equation}

We assume that only a fraction of the flux $F_c$ is generated inside the 
disc due to the viscous forces and the disc carries away only the same fraction
of angular momentum flux (local viscosity model). We therefore follow the 
main line of reasoning underlying
the accreting corona models discussed by Nakamura \& Osaki (1993), 
\. Zycki et al. (1995) and Witt et al. 
(1997). This assumption is not a unique possibility. Some authors assumed 
that the disc carries the entire angular momentum flux even if only a 
fraction of energy is dissipated there (global viscosity model) 
(e.g. Svensson \& Zdziarski 1994,
\. Zycki et al. 1995, Sincell \& Krolik 1997). We discuss this problem in 
Section 3.

The  influence of hot corona appears through the boundary condition. One of the 
effects is the illumination of the disc surface by X-ray radiation 
flux generated in the corona. 
We assume that half of the flux  $F_c$ is directed toward the 
disc and all of it  is 
absorbed  (albedo equals zero) by the disc matter 
and than reemited with addition to the flux dissipated  
in the main  body of the disc:
\begin{equation}
{dF \over dz} = {3\over 2} \alpha P \Omega_K+{1\over 2}F_c \kappa \rho \exp(-\tau). 
\end{equation}

The opacity $\kappa$ (the Rosseland mean) as a function of density and 
temperature is taken from Alexander, Johnson \& Rypma (1994) 
for log $T <3.8$, from Seaton et al. (1994) for  log $T >4.0$ and it is 
interpolated between these two tables for intermediate values of the
temperature. 
                
In the present paper we use the same description for opacity of the incident
X-ray photons and of diffusing UV photons. It means an error up to a factor
of few since the absorption of hard X-ray photons at $\sim 100$ keV can be 
approximately described by opacity $\sim 0.5$ of the scattering opacity 
while in soft X-rays the absorption of the photons by a matter of the 
typical considered ionization stage is of order of $10$ times larger
than the Thompson cross section. Other authors describing the X-ray heating
also use some kind of approximation e.g. Sincell \& Krolik (1997).
Full treatment of this problem including hydrostatic equilibrium is still 
to be done and we will address it in the future. 

The role of convection is important even if it carries not more that 
$\sim 30 \%$ of  the flux.
Here we adopt a simple description of the convection based on 
mixing length theory
used in stellar interiors. This method incorporates the radiation pressure
gradient in the optically thick gas but it is not appropriate in the 
optically thin surface layers (Smak 1982a, 
Meyer \& Meyer-Hofmeister 1981,1982). Actual 
calculations of the disc structure are done using the code described 
by Pojma\'{n}ski (1986)
and applied to AGN by Siemiginowska, Czerny \& Kostyunin (1996);
thermodynamic functions are calculated as in Paczy\'{n}ski (1969).
The effect of partial ionization of the gas whilst calculating the mean 
molecular weight, $\mu$, is included (whenever necessary). 


Equations (7) -- (11)  are integrated from the top of the disc, assumed to be
at $z=H_d$,
toward the equatorial plane, with the initial values for $F$, $T$, and $\rho$
given by the boundary conditions at the bottom of the corona.

At the equatorial plane we require:
\begin{equation}
F(z=0) = 0.
\end{equation}

This condition actually determines $H_d$. 
We solve this two point boundary problem by a shooting method. 
The integration
is performed by the second order Runge--Kutta scheme with adaptive stepsize.
In the absence of the corona ($f=0$) the required
gas pressure at the surface is zero, i.e., consistently, there is no
dynamical influence of the corona on the disc structure.

\subsection{The vertical structure of the corona}

In the most general case the influence of the corona on the disc structure is
given by the coronal X-ray radiation flux (i.e. the fraction of energy generated
in the corona)  and the pressure at the bottom 
of the corona. In order to reduce
these two arbitrary parameters to a single arbitrary parameter we make the
following assumptions about the corona.

We assume that the corona is in hydrostatic equilibrium so the pressure at the
bottom is given by 
\begin{equation}
P = {2 \over \pi} \rho \Omega_K^2 H_c^2,
\end{equation}
where $H_c$ is the scale high of the corona.

We describe the thermal balance in the corona as in the paper of Shapiro, Lightman
\& Eardley (1976) in the version applied in 
the coronal paper of Haardt \& Maraschi (1991), 
i.e. the corona is two-temperature, with heating of ions,
Coulomb transfer of the energy from ions to electrons and Compton cooling 
of electrons:
\begin{equation}
P= {k \over m_H} \rho T_i,
\end{equation}
\begin{equation}
F_c=A F_{soft} ; \\ F_{soft} = 0.5 F_c + F_d,
\end{equation}
where $A$ is the Comptonization amplification factor.

In the present paper the  Comptonization amplification factor is computed
using our
comptonization Monte Carlo code.. The code is based 
on the method developed
by Pozdnyakov, Sobol \& Sunyaev (1983). We have implemented this method as
described by G\'{o}recki \& Wilczewski (1984). We have assumed slab
geometry and a black body soft photon input and we computed the
amplification factor on a grid of $T_{bb}$, $k T_{e}$ and optical 
depth of corona
$\tau_c$, and then interpolated it to values of interest. 

The dependence of our amplification factor on the coronal optical depth 
is shown  on Fig. \ 1 for different values of $T_{bb}$ and  $k T_{e}$.
We compare our values with two analytical approximations for $A$.
The first approximation (dotted line) was given by Dermer, Liang \& Canfield 
(1991), and second
one (dashed line) is $A=e^{y}-1$, where the 
Compton parameter in the corona is: $ y=\kappa_{es} \rho H_c (\kappa_{es} 
\rho H_c+1){4 k T_e /m_e c^2}\left( 1  + {4 k T_e/ 
m_e c^2} \right)$.

All slopes do not depend significantly on the temperature of 
soft radiation from
the disc $T_{bb}$. They are similar even for the case of an accretion discs in
AGNs ($T_{bb} \sim 4$eV) and for accretion discs around galactic black holes
($T_{bb} \sim 100$eV). 

The formula of Dermer et al. (1991) differs from our Monte Carlo values, mainly
because our input photon spectrum is a blackbody whilst they considered
monoenergetic input. Also, their most important parameter is the effective
optical depth which has to be fitted for a given soft photons input energy.

The second analytical approximation is fairly accurate for low electron 
temperatures and optical depths i.e.\ when $y<1$. Although for small $\tau_c$
and high $T_e$ it becomes inaccurate due to non-diffusive character of
Compton scattering in that regime.

The cooling of ions in the corona  by the electron--ion Coulomb 
interaction is described by the following equation (Shapiro,
Lightman \& Eardley 1976)
\begin{equation}
F_c ={3\over 2}  {k (T_i -T_e) \over m_H}
\left[ 1 + \left( {4 k T_e\over m_e c^2}\right)^{1/2} \right]
\int\limits_{z_0}^\infty \nu_{ei} \rho\, dz,
\end{equation}
where $T_i$, $T_e$ are the ion and electron temperatures, and
\begin{equation} 
\nu_{ei}= 2.44 \times 10^{21} \rho T_e^{-1.5}\, \ln \Lambda ~~~[(\rm s)^{-1}];
\quad {\rm with}  \\ \ln \Lambda \approx 20  
\end{equation}
is the electron-ion coupling rate.

Finally, we assume that the bottom of the corona is defined by achievement 
of the thermal instability of the irradiated gas and a rapid 
switch from Compton
cooling to atomic cooling, as described by  Krolik et al. (1981) and
applied to the disc/corona transition by R\' o\. za\' nska (1997). 
The ionization
parameter $\Xi$ 
\begin{equation}
\Xi = {F \over c P_{\rm gas}} 
\end{equation}
is fixed at the bottom of the corona by the scaling properties of its value with the temperature
(Begelman, McKee and Shields 1983)
\begin{equation}
\Xi_b = 0.65 (T_e/10^8)^{-3/2} 
\end{equation}

The structure of the corona is therefore uniquely described by the coronal
flux $F_c$, or a 
fraction $f$
of the energy liberated in the corona for a given accretion rate. 
The equations are equivalent to 
equations used by (Witt, Czerny \& \. Zycki 1997) in Appendix D apart from
their equation (D4) which requires an additional assumption of viscous energy generation
in the corona, not used in the present paper. 

\begin{figure}
\epsfxsize = 100 mm \epsfbox[30 370 490 700]{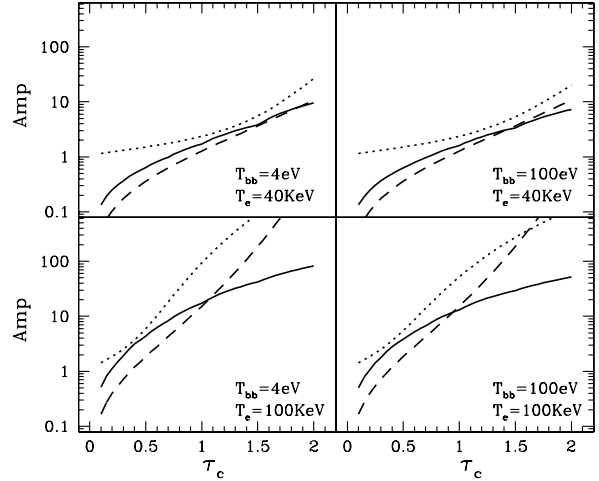}
\caption{Amplification factors for Comptonization versus optical depth of
corona computed in three different methods:
monte carlo simulations - solid line; approximation by Dermer (1991) - 
dotted line; analytical approximation $A=e^{y}-1$ - dashed line.
 $T_e$ is the temperature of 
electrons in corona and $T_{bb}$ is the temperature of soft radiation from the 
disc.
Other parameters are:
radius $r = 10 R_{Schw}$, $M = 10^8 M_{\odot}$, $\dot m = 0.03$}
\end{figure}

\subsection{The boundary conditions between the disc and the corona}

We use the Eddington approximation on the disc surface
and  we assume for $\tau=0$:
\begin{equation}  
F(H_d) = F_{soft},  
\end{equation}
\begin{equation}
\sigma T^4(H_d) = {1\over 2}\sigma T_{eff}^4 ={1\over 2} F_{soft}
\end{equation}

Next boundary condition results from the value of the pressure at the basis of the
corona and the requirement of the hydrostatic equilibrium between the disc and 
the corona 
\begin{equation}
\rho_s ={ F_c \over 2 c \Xi} {\mu m_H \over k T}.
\end{equation}
We take into account only gas pressure since corona generally reduces the
radiation pressure.

\section{Results}

All the numerical results shown are computed at the radius
10$\,R_{\rm Schw}$ being fairly representative for the innermost part of 
the disc responsible for generating most of the energy, 
where $R_{Schw}$, the Schwarzchild radius, equals $2 G M /c^2$.
Central black hole mass and viscosity parameter
in disc and corona are taken: $M=10^8M_{\odot}$,  $\alpha=0.1$. 

To show the vertical profiles of the physical quantities we choose an
accretion rate (in units of the Eddington accretion rate) $\dot m = 0.03 $, 
a value fairly representative for Seyfert
galaxies.

\subsection{Corona properties}

In this paper we do not concentrate on the properties of the corona
itself and the production of hard X-ray emission. This problems are
extensively studied by Witt et al.(1997) and Janiuk \& Czerny (1998)
in the specific context of accreting corona model. 

However, in order to illustrate the basic corona properties which might
be important from the point of view of the conditions on the disc
surface we show the dependence of the coronal optical depth, ion and
electron temperature and the pressure scale height in the corona on the
assumed fraction of energy generated in the corona $f$ (see Fig.\ 2).

\begin{figure}
\epsfxsize = 100 mm \epsfbox[30 370 490 700]{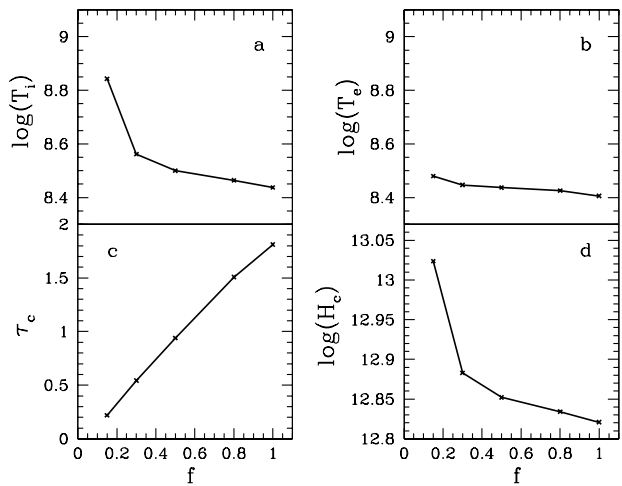}
\caption{Dependence of the coronal a) ion temperature $T_i$, b)
electron temperature $T_e$, c) optical depth $\tau_c$ and d) pressure
scale height $H_{c}$ on the fraction of energy generated in the corona.
Other parameters are:
radius $r = 10 R_{Schw}$, $M = 10^8 M_{\odot}$, $\dot m = 0.03$}
\end{figure}

We see that the ion temperature is generally not much higher than the electron
temperature  in our model  (Fig.\ 2 a,b). This is different from 
other models of corona (Witt et al. 1997) when the equation of energy 
generated in corona is taken into account. We plan to consider such a corona 
above an accretion
disc in the future.  

The optical depth of the corona increases with $f$. The corona is optically
thin for $f\la 0.5$ (Fig.\ 2c).
Since the 
opacity in corona is dominated by electron scattering, the column density in
the corona is of order of $\Sigma_{c}=\tau_{c}/\kappa_{es} \sim 1-5~
{\rm g cm}^{-2}$.

Fig. \ 2d shows that 
under the condition of hydrostatic equilibrium the height of the corona
is of order of $\sim 7\times 10^{12}$ cm, similar to the disc thickness 
(see below).  

\subsection{The effect of the corona weight on the disc surface layers}

The presence of the corona modifies  the conditions close to the
surface of the disc.
In the absence of the corona the density at the disc surface is zero,
consequently the gas close to the surface is fully ionized
and the opacity is reduced to electron scattering. However, even energetically
weak corona imposes certain pressure at the disc surface so the density
at the surface is finite, and it can be quite large (Fig.\ 5b-f).

Let us first consider an atmosphere in an accretion disc without a corona. 
The column density $\Sigma_{ph}$
above the photosphere is of order of $\Sigma_{ph}=\tau_{ph}/\kappa_{es} 
\sim 2$ g cm$^{-2}$ since the opacity in the case is dominated by the electron
scattering. If such a photosphere has the temperature of order of the
effective temperature of the disc $T_{ph} \sim T_{eff} \sim 4 
\times 10^4$ K appropriate
for the parameters adopted in the computations ($M=10^8 M_{\odot}$, 
$r=10 R_{Schw}$, $\alpha=0.1$, $\dot m=0.03$) then the scale height of such a
photosphere is of order of 

\begin{equation}
H_{ph} = ({k T_{ph} \over \mu m_H} {r^3/ G M})^{1/2} \sim 8\times 10^{10} 
 {\rm cm}
\end{equation}
and the pressure imposed by the photosphere is 
\begin{equation}
P_{ph} = {G M \over r^3} \Sigma_{ph} H_{ph} \sim  85~ {\rm erg cm}^{-3} 
\end{equation}
However, if a hot corona develops it has column density 
of the same order as the disc photosphere, but since its ion 
temperature is of order of $10^{9}$ K, its scale height is 
two orders of magnitude higher (Fig. \ 2 d) and the pressure imposed 
on the disc surface is higher by the same factor.
 
This pressure changes the density profile close to the disc surface leading
to more complex behavior than in the disc without the corona. 

The density at
the disc surface is much higher than in the case of a bare disc.
The initial high value of the density decreases towards the
disc interior forming the outer layer of the density inversion (for inner one -- 
see Sec.\ 3.3).
The effect is caused by the dynamical
pressure of the corona. We can see the role of the surface pressure
from an approximate expression

\begin{equation}
\frac {d \rho}{d \tau}=\frac {2}{3} \frac{g}{\kappa_0}-\frac{1}{4} P_0
-\frac{1}{8}a T_{eff}^4 < 0 
\end{equation}
where $P_0$ and $\kappa_0$ are the pressure and opacity coefficient 
at  the disc surface, and $g$ is the surface gravity given by
$GMH_d/r^3$. The region of density inversion is dynamically unstable
which results in the development of convection (Hansen \& Kawaler 1994).
Convection, however, is actually very inefficient in this zone according
to our present computations, in opposite to convection which develops
in the disc interior (see Section~\ref{cor_int}). It remains to be
seen whether this conclusion would change after introducing better
description of the radiative transfer in the disc surface layers since
it may change significantly the temperature gradient close to the
disc surface.
 
The increased value of the density close to the surface 
leads
to a layer of increased opacity (Fig. \ 7). Therefore, whilst for a disc 
model without
a corona the opacity at $\tau = 2/3$ is dominated by the electron scattering,
for a disc with a corona (e.g. for $f = 0.3$)  the 
opacity there is clearly dominated by
bound--free opacities. What is more, the density at $\tau = 2/3$ is also
significantly different in those two cases: it is equal $8.8 \times 10^{-11}
{\rm g cm}^{-3}$ for a bare
disc and $1.24 \times 10^{-9}{\rm g cm}^{-3}$ for the model with  
$f = 0.5$. The expanded plot of the density and opacity
profiles close to the disc surface is shown in Fig.\ 3 and Fig.\ 4.

\begin{figure}
\epsfxsize = 70 mm \epsfbox[30 160 490 700]{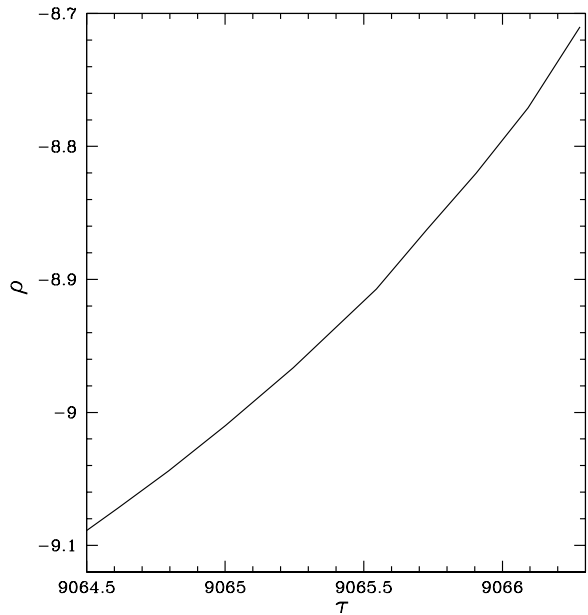}
\caption{Density  profile close to the convective  disc surface for $f=0.5$.}
\end{figure}
\begin{figure}
\epsfxsize = 70 mm \epsfbox[30 160 490 700]{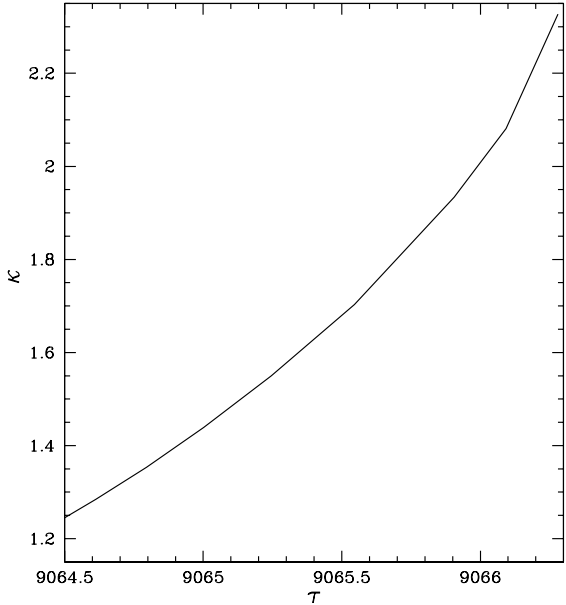}
\caption{Opacity profile close to the convective disc surface for $f=0.5$.}
\end{figure}

Therefore, the density in the region where the disc spectrum 
forms is not much lower than the density at the equatorial plane -- or even
higher for strong coronae. It is more than an order of magnitude lower than
the mean density calculated from the formula of Shakura \& Sunyaev (1973)
\begin{equation}
n^{SS} = 8.3 \times 10^{16} \alpha^{-1} \dot m^{-2} m^{-1} x^{3/2} f(x)^{-2}
\end{equation}
where $x = r/R_{Schw}$ and  $m= M/10^8 M_{\odot}$ which gives the value $n=
2 \times 10^{15}\ {\rm cm^{-2}}$ (or $\rho = 3.35 \times 10^{-9} 
{\rm g\ cm}^{-3}$) for the model shown in Fig.\ 3.

The density at the disc surface, $\rho_s$, depends only weakly
on accretion rate 
as it is shown in Fig.\ 9 for  $f=0.3$ and $f=0.8$.
Since the effective temperature increases with increasing $\dot m$,
the density at $\tau=2/3$ decreases with increasing $\dot m$.

\subsection{The effect of the corona on the disc interior}
\label{cor_int}

The results presented in this section depend significantly on the assumption
that the disc carries only the fraction of the angular momentum proportional 
to the fraction of the energy generated there. Therefore, if e.g.\ almost all
energy is released in the corona ($f \approx 1$) the disc becomes almost 
isothermal
(since it is only heated from outside by the incident X-ray flux), 
geometrically very thin and dense. Its surface density $\Sigma $ 
is slightly lower than in the case of a disc without a corona. 

\begin{figure}
\epsfxsize = 100 mm \epsfbox[30 160 490 700]{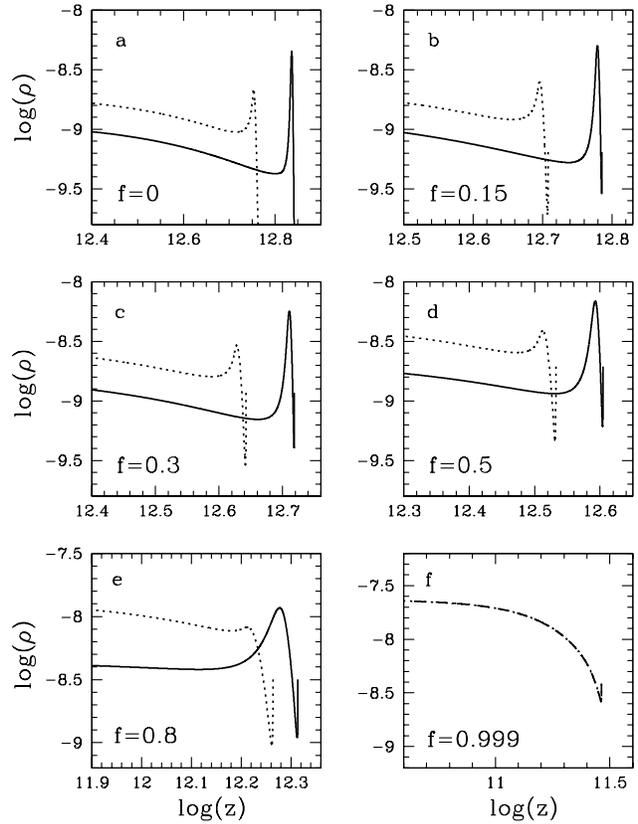}
\caption{The density profiles in the accretion disc with convection (dotted
line)  and without convection (solid line) for a) $f=0$ -  bare disc;  b)
$f=0.1$ -  disc with weak corona; c) $f=0.3$, d) $f=0.5$ - disc with moderate 
corona; e) $f=0.8$ disc with strong corona. In the case when almost all energy is 
dissipated in corona - f) $f=0.999$, it is no difference between disc with
convection (dotted line) and without convection (short dashed line).
Other parameters are:
radius $r = 10 R_{Schw}$, $M = 10^8 M_{\odot}$, $\dot m = 0.03$}
\end{figure}

The profiles of temperature and density in the vertical
direction are very flat throughout most of the disc interior (Fig.\ 5
and Fig.\ 6). 
Their values change rapidly only near
the surface. 

A noticeable feature is the  density inversion at the optical depth of the
order of $100$ (inner density inversion)  which occur irrespectively
of the presence or absence of the corona until the corona becomes very strong.
The possibility of this inversion
can be seen
from the density gradient expression,
\begin{equation}
{d\rho\over dz}=\left[{dP\over dz}-\left({\partial P\over\partial T}
\right)_{\rho}{dT\over dz}\right]\bigg/\left({\partial P\over\partial\rho}
\right)_T,
\end{equation}
which allows for  inversion if the temperature gradient is large enough.
 
Such a density inversion is well known in some classes of stars, like
supergiants (see e.g. Heger \& Langer 1998 for a recent study of supergiant
envelopes). This region is dynamically unstable which results in the
development of convection (Hansen \& Kawaler 1994) which has to be 
included in the computation
of the stellar structure throughout this region although
heat transport by convection is not always efficient. The effect is
usually caused either by distribution of energy sources or by opacity
in partial ionization zone.

Similar situation takes place in the accretion disc interior. 
Radiation pressure dominates throughout the bare disc 
($\beta\equiv P_{\rm gas}/(P_{\rm gas}+P_{\rm rad})\lse 0.01$)
for $\dot m\gse 0.03$. 
The energy flux
to be transported rises approximately linearly with the distance from the 
equatorial plane, therefore, further out from the equatorial plane, 
where pure radiative transfer would cause difficulties, 
a fraction  of the energy
is carried out by convection. This is demonstrated in Fig.\ 5  where we
plot density profiles in the case with and without convection for different 
fractions  of energy dissipated in corona $f$. 
Fig.\ 5a-e show that convection dominates in the disc interior and,  
modifying  the density profile,  reduces  the inner density inversion.
These results are in agreement with calculations done by  Milson et al. (1994)
for accretion disc around low mass black holes and neutron stars. For 
the black hole of $10 M_{\odot}$ 
convection smoothes out inversion almost completely (Milson et al. 1994
Fig.\ 2b).
Comparing the gradients: adiabatic, radiative and the gradient 
$\nabla=d\log T/d\log\rho$ actually achieved in the disc interior
we calculated that the convection carries significant fraction of the flux
($\sim 30\%$).

Another phenomenon which reduces inner density  inversion is the presence of a 
corona.  Note that for bare disc the inversion is strongest.
Stronger corona leads  to lower inversion and for $f=0.999$ this effect  
disappears completely.
When almost all energy is dissipated in the 
corona (Fig.\ 5f) the disc 
is squeezed to the thin and dense slab with purely radiative energy transport. 

\begin{figure}
\epsfxsize = 70 mm \epsfbox[30 160 490 700]{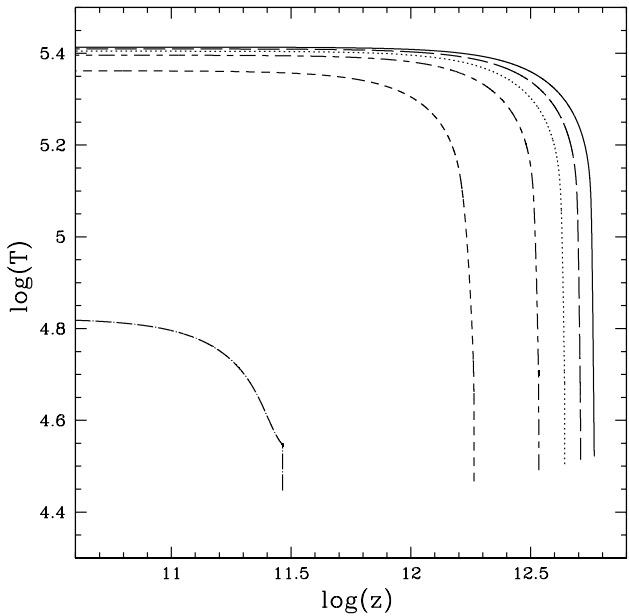}
\caption{The temperature profile in the convective accretion disc without corona
($f=0$; continuous line), with weak corona ($f=0.1$; long dashed line)
with moderate corona ($f=0.3$; dotted line),($f=0.5$; short dashed long dashed
line), with strong corona ($f=0.8$; short dashed line),
with corona which dissipates all energy ($f=1$; dashed dotted line).
Other parameters are:
radius $r=10 R_{Schw}$, $M=10^8 M_{\odot}$, $\dot m = 0.03 $}
\end{figure}

\begin{figure}
\epsfxsize = 70 mm \epsfbox[30 160 490 700]{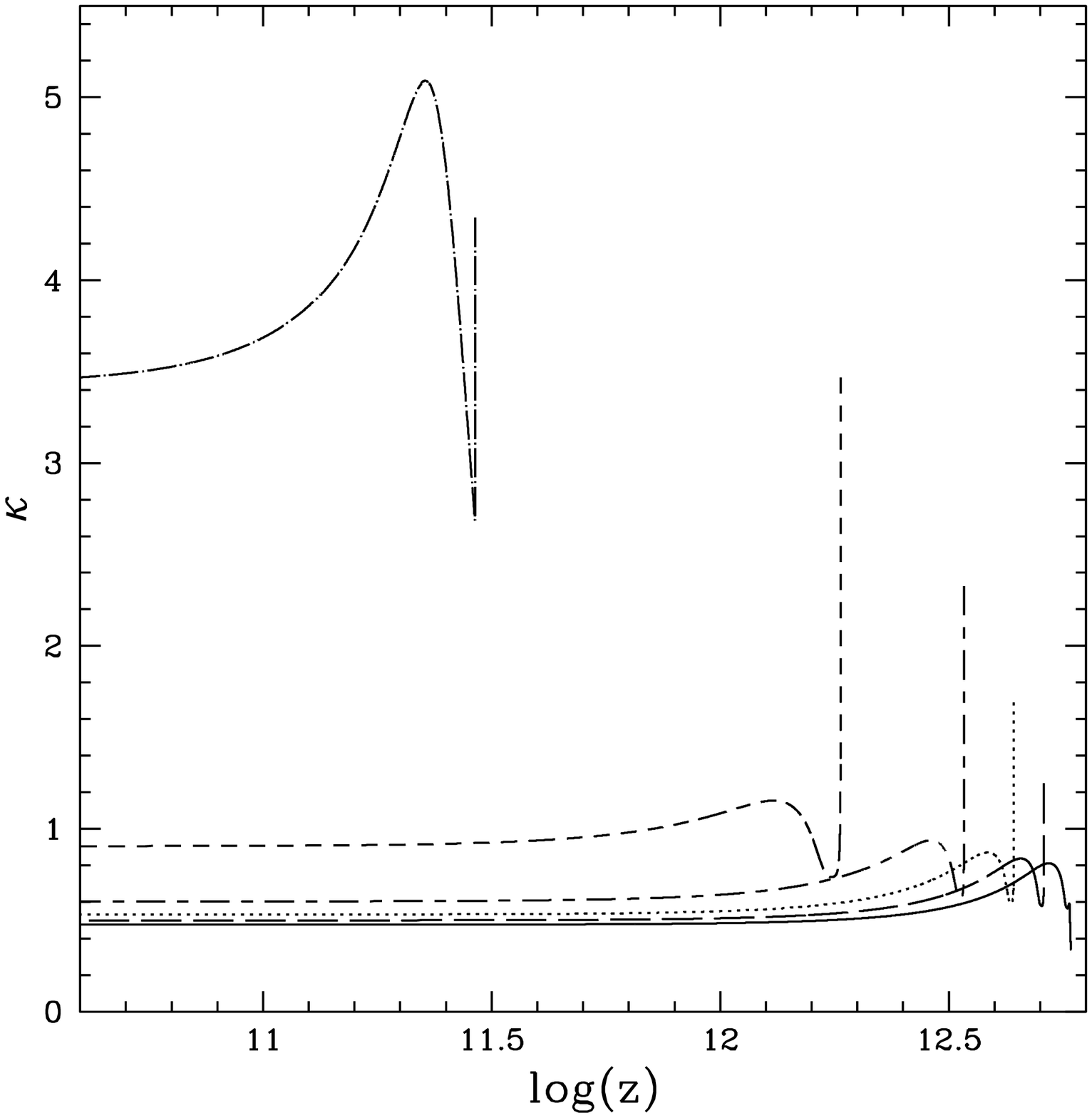}
\caption{The profile of the opacity in the convective accretion disc without corona
($f=0.1$; long dashed line), with weak corona ($f=0.1$; long dashed line)
with moderate corona ($f=0.3$; doted line), ($f=0.5$; short dashed long dashed
line), with strong corona ($f=0.8$; short dashed line),
with corona which dissipates all energy ($f=1$; dashed dotted line).
Other parameters are:
radius $r=10 R_{Schw}$, $M=10^8 M_{\odot}$, $\dot m=0.03 $}
\end{figure}

\subsection{The comparison with global viscosity model}

The results are quite different if the disc has to carry the total
angular momentum flux. In such case strong viscous stress has to operate
within the cold disc even if the energy is generated entirely in the corona.
Numerically, such computations correspond to the change of the surface 
boundary condition for equation (11) from $F_d + F_{\rm X}$ (where $F_{\rm X}$
is the incident X-ray flux)to $F_d + F_c + F_{\rm X}$.

The pressure in the disc interior has now to be high in order to provide 
the integrated stress high enough to transport the angular momentum,
even if the disc does not dissipate the energy.
Therefore the disc column density for e.g. $f=0.999$ is much higher than 
under the assumptions adopted previously.

\begin{figure}
\epsfxsize = 100 mm \epsfbox[30 160 490 700]{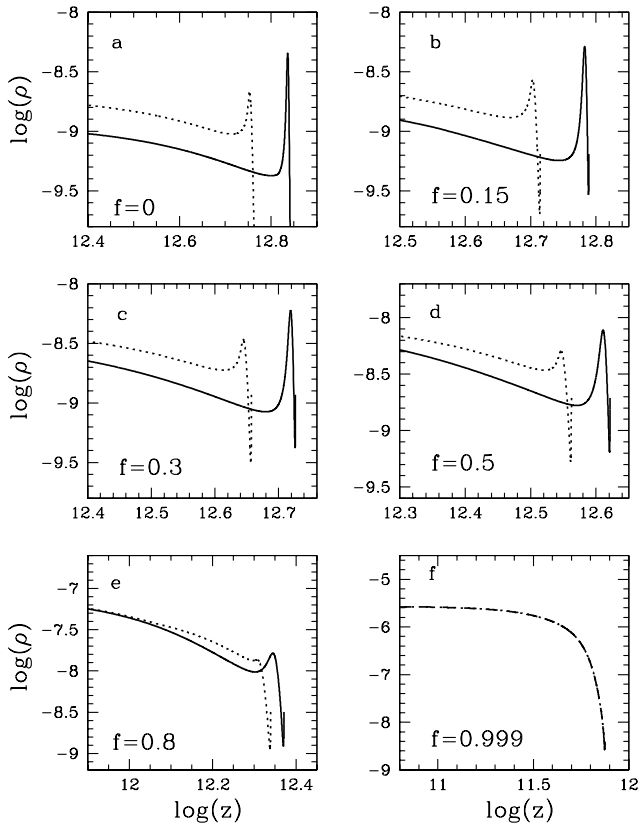}
\caption{The density profiles for global viscosity model of the accretion 
disc with convection (dotted
line)  and without convection (solid line) for a) $f=0$ -  bare disc;  b)
$f=0.1$ -  disc with weak corona; c) $f=0.3$, d) $f=0.5$ - disc with moderate 
corona; e) $f=0.8$ disc with strong corona. In the case when almost all 
energy is 
dissipated in corona - f) $f=0.999$, it is no difference between disc with
convection (dotted line) and without convection (short dashed line).
Other parameters are:
radius $r = 10 R_{Schw}$, $M = 10^8 M_{\odot}$, $\dot m = 0.03$}
\end{figure}

We show examples of solutions for the density profiles in Fig.\ 8 for different
values of $f$ (the same as in Fig.\ 5 for local viscosity model).
We see that for $f$ close to 1 the density in the disc interior is much 
higher than in the 
local viscosity model, keeping the same structure near the surface.

For global viscosity model, in case where almost all energy is 
dissipated in corona ($f=0.999$) the column density of the disc is 
higher of the  order of two magnitude 
than in our previous approach (Fig.5 f and Fig.8 f). This is not surprising
since global viscosity model requires the disc to transport the entire
angular momentum while in case of corona transporting the angular momentum
as well as dissipating the energy, the role of the disc is reduced to
reprocessing half of the X-ray emission.   
   
\section{Lyman edge and Ionization state of the disc matter}

In this paper we do not calculate the spectrum of the disc with the corona 
since further developments of the model are necessary before we
can do that reliably (see Discussion). However we have shown above  that 
the corona 
modifies the layer where  spectrum is formed,  making this layer    
denser and dominated by bound--free absorption.
In this situation the electron scattering is not important 
any more and we can 
expect not to see Lyman edge in emission.  Such an  emission edge is 
predicted in 
all  models computed for bare but irradiated discs (Sincell \& Krolik 1997),
 but it does not appear in any observed 
spectrum. 

Simple estimates  of the dependence of Lyman edge magnitude on effective 
temperature
and density of emitting layer   
were done by Czerny \& Pojma\'nski (1990). They shown the reduction of 
Lyman edge for densities higher than  $10^{-7} {\rm gcm^{-3}}$ and for
wide range of  
effective temperatures higher than $5\times 10^4$K. 
Trying to apply this prediction to the accretion disc model 
they concluded that for bare disc the Lyman edge size close to zero 
corresponds to a very narrow range of accretion rates.  

In the our case of accretion discs with hot coronae  we assumed Eddington 
approximation as in Czerny \& Pojma\'nski (1990).
Making additional assumptions that the expected spectrum is a modified 
blackbody and the disc inclination angle is $i=0^{\circ}$ we
used the same code 
to calculate the Lyman edge for 
various  densities and effective temperatures resulting from our model.

The computations for several values of  accretion rate $\dot m=0.02 - 
0.09 $ and $f$ are presented in Table 1.
We  can see that for bare discs a strong edge in emission is obtained 
for all $\dot m$ but for each $\dot m$ there is  
range of $f$ such that the edge is very weak.

For $\dot m < 0.02$ the edge is in absorption, independently of the strength
of the corona. Larger values of accretion rate allow for an edge both
in absorption and in emission, depending on the strength of the corona.
For increasing accretion rates the range of $f$ for which the edge 
is weak moves towards strong coronae. 
When the corona is relatively strong, the surface density $\rho_s$ 
is high  and an edge in 
absorption is expected if $\dot m \la 0.06$.  Larger accretion rates
are characretised by an edge in emission.
However, for very high accretion rate ($\dot m =0.09$) strong corona cannot 
exist in hydrostatic equilibrium and we cannot obtain solutions 
determining the disc/corona structure.

The presence of corona leads to a wide range of accretion rates 
for which the Lyman edge is significantly reduced in the spectrum.

\begin{table} 
\caption{The size of Lyman edge in magnitudes $\Delta m$ 
          for various  $f$ and $\dot m$. $\Delta m < 0$ indicates the edge
         in emission.}

\begin{tabular}{||c|c|c|c|c|c||}
                          \hline
$f$  & $\dot m=0.02 $ &  $\dot m=0.03$ & $\dot m=0.06$ & $\dot m=0.09$ \\
                           \hline
0  & -0.7 &  -0.9   & -1.07 & -1.13 \\
                   \hline
0.15 & -0.14 & -0.37   & -0.63  & -0.76 \\
                  \hline 
0.3 & -0.01 & -0.21  & -0.43 & -0.56  \\
                  \hline 
0.5 & 0.26 &  0.02  & -0.26 & -0.40 \\
                  \hline
0.8 & 0.58 & 0.36  & -0.07 & -0.16 \\
                   \hline 
0.999 & 0.74& 0.56  & 0.14 & ---  \\
                   \hline 
\end{tabular} 
\end{table}

Trying to predict some spectral features of our model, we 
can compare the estimate 
of the ionization parameter $\xi$ at the disc/corona surface with the
limits derived from from the properties of the reflection component
This parameter is usually defined as 
\begin{equation}
\xi = {4 \pi F \over n}
\end{equation}
where $F$ is the incident flux (equal $0.5 F_c$ in our model) and $n$ is the
number density of the irradiated layer (note the essential difference from
ionization parameter $\Xi$ used in equation 19). 
For the case shown in Fig.\ 3 ($f = 0.5$) the value of this parameter 
at the optical depth $\tau = 2/3$ is
equal 0.3 whilst for a disc with the same parameters but without corona,
this value is equal 100.

The properties of the interior change dramatically only if almost all the
energy is liberated in the corona ($f > 0.999$ ) and the disc only 
reprocesses the X-ray radiation. In that case the disc becomes almost 
isothermal and dominated by the gas pressure. 
The mean density 
of the interior is not monotonic as a function of $\xi$: the largest density 
is obtained for intermediate $\xi$. 
This is connected with $P_{gas}/P_{rad}$
ratio, the maximum $\rho$ being obtained when both pressures
are comparable. 

\begin{figure}
\epsfxsize = 70 mm \epsfbox[30 160 490 700]{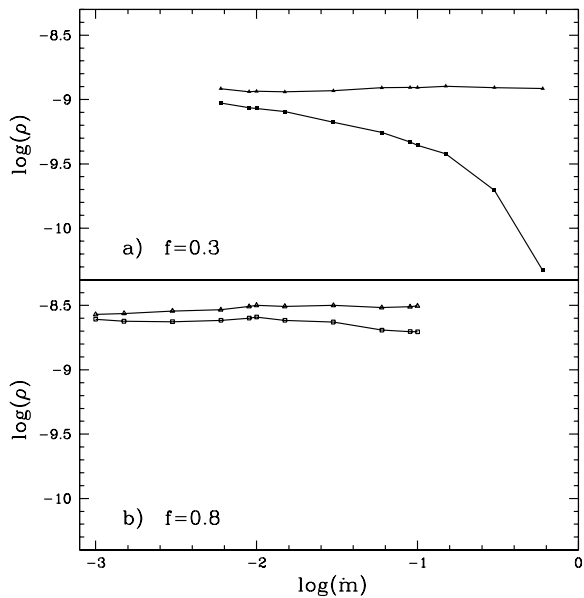}
\caption{The dependence  of surface density (triangles), and the density of
the layer on $\tau=2/3$ (squares), on accretion rate
for two cases of corona a) $f=0.3$ and b) $f=0.8$.
Other parameters are:
radius $r=10 R_{Schw}$, $M=10^8 M_{\odot}$.}
\end{figure}

\section{Discussion}
   
Previous computations of the disc spectra were done by a number of authors
(e.g. Shimura \& Takahara 1993, Ross, Fabian \& Mineshige 1993) but none of
these papers treated carefully the disc surface where the radiation spectrum 
forms so these spectra models are not ready yet for detailed comparison with
the observational data.

In this paper we describe the vertical structure of the disc instead of
solving algebraic equations for the vertically averaged quantities.
It allows us to incorporate properly the effect of the corona, 
which modifies
significantly the outer layers of the disc.
A hot corona exerts a dynamically pressure on the disc and 
changes its surface zone making it denser and
dominated by bound -- free absorption.  The first  consequence  of the 
coronal  influence is  the low ionization state of  the matter on
the disc surface. 
This agrees with observations for which we know that X-ray 
radiation produced in 
corona reflects from the relatively neutral gas $\xi \sim $ few 
(Pounds et al.1990,
Nandra et al. 1994).

The size of Lyman edge close to zero for wide range of accretion rates 
is the  second important consequence of our model.  
In the observed quasar spectra we do not see Lyman edge, but all 
bare discs models predict such a feature. Also,  recent calculations
by Sincell \& Krolik (1997) of X-ray heated disc, 
indicate the presence
of emission Lyman edge in the disc spectrum. They considered a layer
resulting from X-ray illumination positioned on top of a cold disc.
Illumination of plasma by relatively hard X-rays leads to existence of
multi--phase medium due to thermal instability (Krolik et al. 1981).
If such a situation takes place in accretion discs, then, as shown by
R\'o\.za\'nska (1998), the plasma in the transition 
layer  between cold disc and X-ray source forms a two--phase medium.
Sincell \& Krolik (1997) considered only one, cold phase of the transition 
layer. Stopping computations just before the temperature runaway they obtained 
the density of the layer which is one order of magnitude lower then 
density at the disc surface in our model. Also, the temperature they compute
is higher, so they could expect Lyman edge in emission. 
However, this treatment can overestimate Lyman edge in emission since it 
does not take 
into account that 
transition layer is a mixture of cold and hot phase. 
In the present approach we also neglected the presence of the two--phase 
medium but we replaced the transition zone by a hot, instead of a cold 
phase which is more appropriate but may lead to some underestimate of 
the edge in emission.
    
However, a number of further improvements to the model are necessary before
we can reliably calculate the spectrum emitted by such a disc/corona model.

First of all, more careful description of the disc/corona transition is
necessary. A fraction of
that zone consists of two-phase medium, possibly undergoing continuous,
quasi-stationary evolution (R\'o\.za\'nska 1998).  
Future models should incorporate this 
transition zone as its optical depth and the temperature profile may have
an important effect on the emergent radiation flux.

Further work is needed to improve the description of the X-ray heated
upper layers of the disc. R\' o\. za\' nska (1998) showed how important
it is in the transition zone. However, her method to iterate between
the vertical structure code and Monte Carlo computations of X-ray transfer
cannot be repeated directly in the present work since those computations 
were based on radiative heating and cooling appropriate only for optically
thin zone. In order to describe the disc we need a code which solves the
radiative transfer in optically thick irradiated medium. Such a code
was developed by Collin-Souffrin \& Dumont (see Collin-Souffrin et al. 1996,
Czerny \& Dumont 1998), and we intend to combine it with our 
description of the disc structure. 

A number of improvements can be done for the disc interior as well, although 
they may not be essential from the point of view of the emitted spectra.

The accretion pattern is more complex then usually considered. 
Two-dimensional studies of models based on local viscosity $\alpha$
(Urpin 1984, Siemiginowska 1988,
Kley \& Lin 1992, R\' o\. zyczka, Bodenheimer \& Bell 1994) showed that
for low viscosity there is actually an outflow near equatorial plane and 
inflow close to the surface;  pure inflow is only for high viscosity 
($\alpha > 0.1$). 

The convection  here 
is described using the standard mixing length theory (with the 
phenomenological coefficient,
also named $\alpha$, equal 1). More advanced approach may require correct
adjustment of this value (see Cannizzo 1992 for discussion of the 
convection 
in the  outer, gas dominated, parts of the disc).

We note an advantage of solving the vertical structure of the disc.
Frequently, the use of vertically averaged algebraic equations for disc
structure leads to  unphysical multiple solutions if complex description 
of opacities is adopted 
as well as a contribution to the pressure both from gas and radiation is
allowed (Cannizzo \& Reiff 1992; Hur\'{e} et al. 1994ab); 
these multiple solutions vanish if 
vertical structure of the disc is integrated step by step
and, for a given $\dot m$, $r$ and $\alpha$, the structure is unique. 
The uniqueness of solutions is preserved in the case of 
a disc with a corona.

Finally, as suggested by the stability analysis of the radiation dominated 
disc and clearly seen in observations, the disc undergoes continuous 
evolution
and the assumption of stationarity can be satisfied only in the sense of
time averaged quantities, at best. Recent computations of disc evolution for
the case of galactic black holes (Szuszkiewicz \& Miller 1998) give the 
timescales $\sim 780 s$ coinciding nicely with the peak in the power density
spectra of these objects (see van der Klis 1995). In the case of AGN only
the slow evolution of the outermost, gas dominated, parts of the discs was
properly addressed (see e.g. Siemiginowska, Czerny \& Kostyunin 1996 and the
references therein).

\section{Conclusions}

In this paper we show that the presence of the corona changes essentially the
physical conditions close to the disc surface. The disc is much denser there
and less ionized than a bare disc due to the dynamical pressure of the hot 
medium. Although at present we do not show yet the radiation spectrum of
such a model we expect that the enhanced surface density makes this disc model
promising. The reflection from such a disc will not show any signs of
high ionization. Also the Lyman edge will be significantly reduced 
in such a model as the
spectrum emitted locally approaches a black body for increasing density. It
will remove the basic argument against an existence of an accretion disc in
AGN based on the absence of the  Lyman edge in the data.
Therefore the corona is not only essential from the point of view
of X-ray spectra formation but helps to remove the problems with discy
models for AGN.


\section*{Acknowledgements}

We thank J. Smak, Z. Loska and W. Dziembowski for helpful discussions, and the anonymous 
referee for remarks leading to significant improvement of the paper and 
clarification of the presented results.
This work was supported in part 
by grant no. 2P03D00410 of the Polish State Committee for 
Scientific Research.

\ \\
This paper has been processed by the authors using the Blackwell
Scientific Publications \LaTeX\  style file.

\end{document}